\documentclass[11pt,reqno]{amsart}
\usepackage[T1]{fontenc}
\usepackage[utf8]{inputenc}
\usepackage{lmodern}
\usepackage[margin=1in]{geometry}
\usepackage{amsmath,amssymb,graphicx,booktabs,longtable,array}
\usepackage{microtype}
\usepackage[authoryear,round]{natbib}
\usepackage{url}
\usepackage[hidelinks]{hyperref}
\usepackage[section]{placeins}
\newcounter{algorithm}
\title[A randomized whole-model test for SVEM]{A Randomized Permutation Whole-Model Test Heuristic for Self-Validated Ensemble Models (SVEM)}
\author{Andrew T. Karl}
\address{Adsurgo LLC, 1141 N Loop E Suite 105 Unit 463, San Antonio, TX 78232}
\email{andrew.karl@adsurgo.com}
\date{March 13, 2024; personal \LaTeX{} rendering revised October 22, 2026}
\thanks{Author's personal rendering of the accepted manuscript, incorporating typographical and other minor corrections. The definitive version appeared in \emph{Chemometrics and Intelligent Laboratory Systems} \textbf{249} (2024), 105122, \url{https://doi.org/10.1016/j.chemolab.2024.105122}.}
\hypersetup{pdftitle={A Randomized Permutation Whole-Model Test Heuristic for Self-Validated Ensemble Models (SVEM)},pdfauthor={Andrew T. Karl}}
\begin{document}
\begin{abstract}
We introduce a heuristic to test the significance of fit of Self-Validated Ensemble Models (SVEM) against the null hypothesis of a constant response. A SVEM model averages predictions from \emph{nBoot} fits of a model, applied to fractionally weighted bootstraps of the target dataset. It tunes each fit on a validation copy of the training data, utilizing anti-correlated weights for training and validation. The proposed test computes SVEM predictions centered by the response column mean and normalized by the ensemble variability at each of \emph{nPoint} points spaced throughout the factor space. A reference distribution is constructed by refitting the SVEM model to \emph{nPerm} randomized permutations of the response column and recording the corresponding standardized predictions at the \emph{nPoint} points. A reduced-rank singular value decomposition applied to the centered and scaled \emph{nPerm} $\times$ \emph{nPoint} reference matrix is used to calculate the Mahalanobis distance for each of the \emph{nPerm} permutation results as well as the jackknife (holdout) Mahalanobis distance of the original response column. The process is repeated independently for each response in the experiment, producing a joint graphical summary. We present a simulation-driven power analysis and discuss limitations of the test relating to model flexibility and design adequacy. The simulations assess Type I error and power; the heuristic can also be applied when the candidate SVEM model contains more parameters than observations.
\end{abstract}
\maketitle

\section{Introduction}\label{introduction}

The Self-Validated Ensemble Model (SVEM) framework \citep{gotwalt2018validation,lemkus2021svem} offers a bootstrap approach to improve predictions from a variety of base learning models such as forward selection regression, Lasso regression, neural networks, and any other statistical or machine learning predictor that is capable of training and tuning on distinct data sets while utilizing fractional weights for the observations. It is particularly suited for situations where a complex response surface is modeled with relatively few experimental runs and has shown promise in many applications \citep{ramsey2021innovation,ramsey2021paradigm,ramsey2023prediction}, including in mixture formulation optimization \citep{ramsey2018validation,karl2023lnp}. The formulation setting is especially challenging due to special accommodations traditionally needed to fit a linear regression model with a mixture constraint and due to high-order interactions that may be present \citep{cornell2002mixtures}. \citet{karl2022mixtures} present simulations demonstrating that SVEM improves formulation optimization performance over several base models. SVEM has potential to be a useful modeling approach for general blending mixture models \citep{brown2015blending}.

As with other regularization methods that allow for fitting models with more parameters than experimental observations, SVEM can produce actual-by-predicted plots with nearly perfect correlation even when the response is independent of the study factors. Ideally, a test set of interior runs will be included in an experiment to serve as a holdout set for the predictive model \citep{ramsey2023prediction}. However, for situations where such a test set is not available, this paper proposes a whole-model test heuristic for the null hypothesis that the response surface is constant over the study factors. While hypothesis testing does not play a primary role in prediction applications and can prove detrimental to their goal \citep{shmueli2010explain}, the proposed test can be useful in multiple response experiments where the optimization uses a joint objective function that consists of a weighted average of the predicted response functions. Scientific knowledge prioritizes the weighting of the responses during joint optimization; however, information about the relative quality of the fit of the model for each response can provide useful information for optimization planning. An insignificant whole-model test result that is contrary to expectations from subject matter knowledge could help detect potential experimental, data recording, or modeling errors.

\begin{figure}[htbp]
\centering
\includegraphics[width=0.82\linewidth]{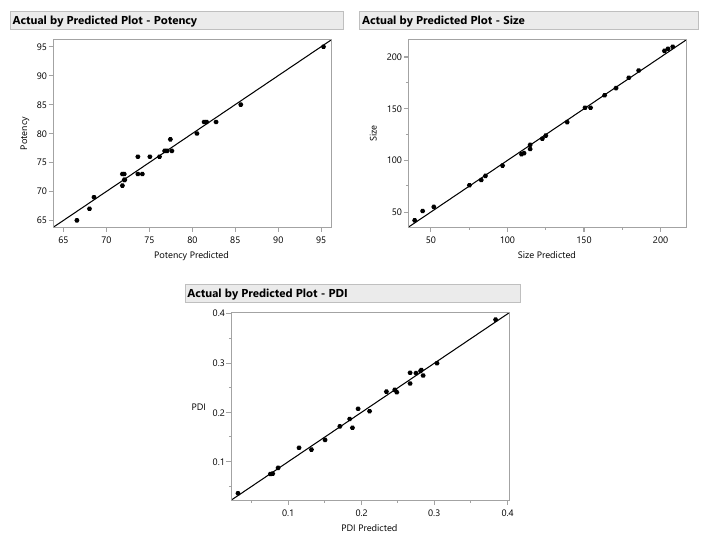}
\caption{Actual-by-predicted plots from the SVEM forward selection models for Potency, Size, and PDI.}\label{_Ref147262584}
\end{figure}

To illustrate, consider a simulated scenario based on a lipid nanoparticle (LNP) mixture-process experiment described by \citet{karl2023lnp}. Suppose a scientist has designed and run an experiment to optimize their formulation composition with respect to three responses: potency (e.g. transfection rate), mean particle size, and polydispersity index (PDI). The experiment consists of 23 runs over four mixture factors, one three-level categorical factor, and two continuous factors. Figure~\ref{_Ref147262584} shows the actual-by-predicted plots from fitting SVEM (with a forward selection regression base model) to the target linear model, which contains more candidate parameters (100) than runs in the experiment. Appendix~\ref{appendix-a} lists the study factors and model effects.

Potency is the highest priority and should be maximized, and Size and PDI are secondary physicochemical responses that should be minimized. In the simulation, Potency and Size depend on the study factors, but PDI is generated independently of the study factors. However, Figure~\ref{_Ref147262584} shows that SVEM produces nearly perfect correlation in the actual-by-predicted plots of all three responses, including PDI. The spurious correlation in the PDI actual-by-predicted plot is a result of fitting 100 candidate parameters with 23 runs: the base forward selection regression model without the SVEM bootstrap produces a similar plot.

In this paper, we introduce a heuristic approach for a test designed to gauge the significance of a fitted SVEM model compared to a null hypothesis of a constant response surface. This method can help identify responses that have relatively stronger or weaker relationships with study factors. The test centers the model predictions by the response mean and divides by the ensemble (bootstrap) standard deviation to create standardized predictions. A reference distribution is constructed for the standardized predictions by repeatedly fitting the same SVEM modeling approach to randomized permutations \citep{edgington2007randomization} of the response vector and recording the resulting standardized predictions.

The section on limitations of the proposed test considers scenarios that could lead to increased error rates. Of particular concern are Type II errors that could lead to a study response being discarded if this test is employed overzealously with a hypothesis testing mindset \citep{shmueli2010explain}. The goal of this paper is to furnish a graphical heuristic as a SVEM diagnostic which can facilitate conversation between researchers. The presented hypothesis testing framework and resulting approximate p-value are used to assess the statistical properties of the graphical summary in different scenarios. The p-value reported by the test for the strength of measured relationship between the response and study factors should be of secondary importance to scientific understanding of the process and employed with an appreciation of the risk of either Type I or Type II errors.

Section~\ref{svem} summarizes the SVEM framework. Section~\ref{proposed-whole-model-test-heuristic} presents the proposed whole-model test algorithm. Section~\ref{illustrative-example} continues the illustrative example from the introduction. Section~\ref{evaluation-of-test-properties} performs a power analysis of the test in three scenarios. Section~\ref{limitations-of-the-methodology} summarizes limitations of the test. Closing comments are provided in Section~\ref{final-comments}.

\section{SVEM}\label{svem}

Building a SVEM framework first requires choosing a base modeling procedure that can use a validation set for tuning. This paper considers both SVEM with a forward selection regression base model (SVEM-FS) and SVEM with a Lasso regression base model (SVEM-Lasso). Both approaches involve specifying a ``full'' regression model which SVEM fits with the base model using different weights for the rows of training data in each iteration, along with anticorrelated weights in the validation data that are used for tuning the base model. An important point here is that reduction from the full model in each iteration, whether with forward selection or Lasso, is done for the utilitarian purpose of improving predictive performance, not for removing ``insignificant'' model effects for the purpose of providing a parsimonious explanatory model \citep{shmueli2010explain}. The final SVEM model averages the predictive models from each iteration and often depends on all of the effects provided as candidates. It is a useful procedure for prediction or optimization but not for identifying statistical significance of individual effects or factors.

In each iteration, SVEM applies a random exponentially distributed weight to each observation. A duplicate dataset is created for validation purposes. The only difference between the training and validation sets is in the weights used for each row in the validation set; these are intentionally selected to be anti-correlated with weights designated in the training dataset for the corresponding observations \citep{gotwalt2018validation,lemkus2021svem,lemkus2021svempreprint}. This is a multivariate analogue of the fractional weight bootstrap \citep{xu2020fractional}. SVEM-FS and SVEM-Lasso fit the regression parameter estimates using the training-weighted response vector while tuning on the validation-weighted response vector.

\citet{lemkus2021svem} recommend \emph{nBoot}=200 SVEM iterations, which we also use in our applications and simulations. This produces \emph{nBoot} prediction vectors, each representing a different model of the response surface. The SVEM prediction for a specific point in the factor space is the average of the corresponding element of the \emph{nBoot} prediction vectors. The ensemble variability in predictions is represented by the standard deviation of the \emph{nBoot} vectors for that specific point and approximates the standard error of the prediction. A low standard deviation of the predictions from all bootstrap sub-models at a point suggests that the base model produces consistent predictions for that point across the reweighted SVEM iterations.

To establish notation for the algorithm: when trained on a continuous response vector \(Y\) and a corresponding design matrix \(\mathbf{X}\), SVEM produces a function \(\mathbf{P}_{\left( \mathbf{X},Y \right)}\left( \mathbf{T} \right)\) where \(\mathbf{T}\) is any matrix of factor levels with the same number of columns (but not necessarily rows) as \(\mathbf{X}\). \(\mathbf{P}_{\left( \mathbf{X},Y \right)}\left( \mathbf{T} \right)\) is an \(\operatorname{nrow}\left( \mathbf{T} \right) \times \ensuremath{\mathit{nBoot}}\) matrix of predictions from the SVEM iterations: the resulting SVEM prediction function is defined as \({\widehat{\mathbf{f}}}_{\left( \mathbf{X},Y \right)}\left( \mathbf{T} \right) = \operatorname{rowMean}\left( \mathbf{P}_{\left( \mathbf{X},Y \right)}\left( \mathbf{T} \right) \right)\). The vector \({\widehat{\mathbf{s}}}_{\left( \mathbf{X},Y \right)}\left( \mathbf{T} \right) = \operatorname{rowStdDev}\left( \mathbf{P}_{\left( \mathbf{X},Y \right)}\left( \mathbf{T} \right) \right)\) provides the standard error of the SVEM prediction at each point. For notational convenience we will use \({\widehat{\mathbf{f}}}_{Y}\) and \({\widehat{\mathbf{s}}}_{Y}\), dropping the reference to the matrix \(\mathbf{X}\) used to train the model where there is no ambiguity.

Because of the variability induced by the random weights employed by SVEM, repeated applications of the same SVEM prediction framework will produce different estimates \({\widehat{\mathbf{f}}}_{\left( \mathbf{X},Y \right)}\). To portray this variability, the heuristic constructs the SVEM prediction vector \emph{nSVEM} (default 5) times for graphical comparison with the reference distribution.

While SVEM is flexible in its ability to model results from a number of different design structures, whether optimal or space filling, the quality of the SVEM predictions depends on how well the experimental design captures regions of complexity in the response surface. If an optimal design targeting a relatively simple response surface model (RSM: main effects, quadratics, two-way interactions) is used, the oversampling of the boundary of the factor space could fail to capture complex behavior beyond what is assumed by the RSM in the interior, resulting in large prediction errors for these regions. In such situations, the proposed test may also fail to detect the relationship between the response and study factors. To protect against this, it is ideal to include a set of interior test runs that can be used to assess the quality of the SVEM fit.

\section{Proposed Whole-model Test Heuristic}\label{proposed-whole-model-test-heuristic}

The proposed test procedure is specified in Algorithm~\ref{alg:whole-model}. First, a description of the approach is provided. Let \(\mathbf{X}\) be a matrix of factor levels. Assume that a continuous response vector \(Y\) is determined by

\[Y = f\left( \mathbf{X} \right) + \epsilon\]
where \(f\) is the true response function and the vector \(\epsilon\) of independent error terms has mean vector 0. For the unrestricted response permutations used below, we additionally assume that the observations are exchangeable under the null hypothesis; identically distributed errors are sufficient in this setting. We wish to test

\[H_{0}:f\left( \mathbf{X} \right) = \mu\mathbf{1}\]

\[H_{A}:f\left( \mathbf{X} \right) \neq \mu\mathbf{1}\]
where \(\mu\) is a constant and \(\mathbf{1}\) is the vector of ones with the same length as \(Y\).

Although SVEM is not restricted to using a linear model (as SVEM-FS and SVEM-Lasso do), it is helpful to recall the structure of the classical ANOVA whole-model test for a multiple linear regression with \emph{n} observations and \emph{p} parameters, which considers the statistic

\[F^{*} = \frac{\mathrm{MSR}}{\mathrm{MSE}} = \frac{\frac{\sum_{i=1}^{n}\left( {\widehat{Y}}_{i} - \overline{Y} \right)^{2}}{p - 1}}{\frac{\mathrm{SSE}}{n - p}}\]
where \(i = 1,\ldots,n\) indexes the observations, \({\widehat{Y}}_{i}\) is the regression model prediction at point \(i\), and \(\mathrm{SSE} = \sum_{i=1}^{n}\left( Y_{i} - {\widehat{Y}}_{i} \right)^{2}\) is the sum of squared errors. This test provides a comparison of the difference of the response surface from the mean (measured at the design points) relative to the global estimate of error variance. However, in situations where \(p \geq n\), the \(F^{*}\) statistic does not follow an \(F\)-distribution due to insufficient degrees of freedom for error estimation: an interpolating fit can give SSE equal to zero, and \(n - p\) will be nonpositive. When \(\mathrm{SSE} = 0\), the actual-by-predicted plot will indicate a perfect fit. Furthermore, when the linear model is tuned on its training data -- as is the case with Lasso or forward selection, with or without SVEM -- this test becomes liberal and, even if \(p < n\), is not valid to apply to the final model. This paper provides a whole-model test heuristic that can be applied when \(p \geq n\) and despite the tuning performed by SVEM on (a reweighted copy of) the training data.

This heuristic considers local comparisons of the response surface to the response mean, not limited to design points (factor settings) used in the experiment. Other work also considers response surface examination via such local comparisons; for example, \citet{stevens2020agreement} present a Bayesian approach for a pointwise comparison of the similarity of two response surfaces with an equivalence testing framework where the probability of the response surfaces falling within a specified equivalence range is calculated over the factor space. In the current application, the response surface is more simply compared to a fixed value.

A prediction for a point \({\widehat{Y}}_{i}\) can be compared to the response mean via the standardized prediction

\[\frac{{\widehat{Y}}_{i} - \overline{Y}}{s.e.\left( {\widehat{Y}}_{i} \right)}\]

When a bootstrap method is employed, the ensemble standard deviation \({\widehat{\mathbf{s}}}_{Y}\) serves as an estimate of the standard error of the prediction. While such standard errors of prediction can have downward bias in traditional bootstrap settings \citep{efron1994bootstrap}, SVEM utilizes the fractional-weight bootstrap \citep{xu2020fractional} -- introduced as the Bayesian bootstrap \citep{rubin1981bayesian} -- and may have different theoretical properties. Rather than deriving the null distribution analytically, our approach uses randomization of the response vector to construct a reference distribution \citep{edgington2007randomization}.

The proposed test heuristic calculates the \emph{standardized prediction}

\begin{equation}\label{eq:standardized}
\frac{{\widehat{f}}_{Y}\left( X_{p_{i}} \right) - \overline{Y}}{{\widehat{s}}_{Y}\left( X_{p_{i}} \right)}
\end{equation}
at each of several locations, \(X_{p_{i}}\), for \(i = 1,\ldots,\ensuremath{\mathit{nPoint}}\). The \(X_{p_{i}}\) are random (or space filling) points throughout the factor space that need not correspond with the points measured during the experiment. \(\ensuremath{\mathit{nPoint}}\) should be large enough to cover the factor space and capture areas where the response surface may deviate from the mean. After the SVEM model has been refit \emph{nSVEM} times, the resulting standardized predictions are recorded in an \(\ensuremath{\mathit{nSVEM}} \times \ensuremath{\mathit{nPoint}}\) matrix \(\mathbf{M}_{Y}\).

The reference distribution for each point \(X_{p_{i}}\) is formed by fitting the SVEM model independently to \emph{nPerm} randomly permuted vectors of the response column, \(\pi_{1}(Y),\ldots,\pi_{\mathit{nPerm}}(Y)\) as shown in Equation~\eqref{_Ref155691832}

\begin{equation}\label{_Ref155691832}
\frac{{\widehat{f}}_{\pi_{j}(Y)}\left( X_{p_{i}} \right) - \overline{Y}}{{\widehat{s}}_{\pi_{j}(Y)}\left( X_{p_{i}} \right)}
\end{equation}

The resulting standardized predictions are recorded in an \(\ensuremath{\mathit{nPerm}} \times \ensuremath{\mathit{nPoint}}\) reference matrix \(\mathbf{M}_{\pi(Y)}\).

A Mahalanobis distance (\(\mathbf{d}_{\pi(Y)}\)) is calculated for the rows of \(\mathbf{M}_{\pi(Y)}\) using a centered and scaled reduced-rank singular value decomposition (SVD) on \(\mathbf{M}_{\pi(Y)}\) \citep{mason2002multivariate}. The median of the jackknife (holdout) Mahalanobis distances (\(\mathbf{d}_{Y}\)) of the rows of \(\mathbf{M}_{Y}\) is compared to the distribution of \(\mathbf{d}_{\pi(Y)}\). The permutation distances provide an approximate reference for this comparison. The holdout distances and their median need not have exactly the same finite-sample distribution as the internally fitted permutation distances; the simulation study assesses the calibration of the resulting heuristic. For a graphical summary, the \(\mathbf{d}_{\pi(Y)}\) and \(\mathbf{d}_{Y}\) are plotted together.

The default settings used by Algorithm~\ref{alg:whole-model} are \(\mathit{nPerm}=125\), \(\mathit{nPoint}=2000\), \(\mathit{nSVEM}=5\), \(\mathit{percent}=85\), and \(\mathit{nBoot}=200\). Ideally, \emph{nBoot} will match the setting employed in the application. A discussion of these settings is provided in Appendix~\ref{appendix-b}. The algorithm uses the notation \(\mathbf{X}[i,\cdot]\) to indicate the \emph{i}-th row of a matrix \(\mathbf{X}\). The supplementary material contains R code implementing the steps of Algorithm~\ref{alg:whole-model} starting with Step 7. The supplementary material also provides a JMP JSL script for the entire algorithm. Algorithm~\ref{alg:whole-model} should be repeated independently for each study response.

\refstepcounter{algorithm}\subsection*{Algorithm \thealgorithm: Whole-model test heuristic}\label{alg:whole-model}
All vector quotients and square roots below are evaluated elementwise. The parameter \(\ensuremath{\mathit{percent}}\) is expressed on a 0--100 scale, with default 85. The ensemble standard deviations and the column standard deviations used for standardization must be nonzero; the retained singular values must be positive. If these conditions fail, the stated calculation is undefined.
\begin{enumerate}
\item Let \(Y\) be a continuous study response vector and \(\mathbf X\) its design matrix. Form \(\mathbf T\) with \(\ensuremath{\mathit{nPoint}}\) random or space-filling points throughout the study factor space. Its number of columns equals \(\operatorname{ncol}(\mathbf X)\).
\item Select a SVEM framework producing the \(\ensuremath{\mathit{nPoint}}\times\ensuremath{\mathit{nBoot}}\) prediction matrix \(\mathbf P_{(\mathbf X,Y)}(\mathbf T)\). Write
\[
 \widehat{\mathbf f}_Y(\mathbf T)=\operatorname{rowMean}\{\mathbf P_{(\mathbf X,Y)}(\mathbf T)\},\qquad
 \widehat{\mathbf s}_Y(\mathbf T)=\operatorname{rowStdDev}\{\mathbf P_{(\mathbf X,Y)}(\mathbf T)\}.
\]
\item Allocate an \(\ensuremath{\mathit{nSVEM}}\times\ensuremath{\mathit{nPoint}}\) matrix \(\mathbf M_Y\).
\item For \(i=1,\ldots,\ensuremath{\mathit{nSVEM}}\):
  \begin{enumerate}
  \item Fit \(\widehat{\mathbf f}_Y(\mathbf T)\) and \(\widehat{\mathbf s}_Y(\mathbf T)\) using a fresh set of SVEM random weights.
  \item Calculate \(\mathbf h_Y=\{\widehat{\mathbf f}_Y(\mathbf T)-\overline Y\}/\widehat{\mathbf s}_Y(\mathbf T)\), subtracting \(\overline Y\) from each component.
  \item Set \(\mathbf M_Y[i,\cdot]\leftarrow\mathbf h_Y^{\prime}\).
  \end{enumerate}
\item Allocate an \(\ensuremath{\mathit{nPerm}}\times\ensuremath{\mathit{nPoint}}\) matrix \(\mathbf M_{\pi(Y)}\).
\item For \(j=1,\ldots,\ensuremath{\mathit{nPerm}}\):
  \begin{enumerate}
  \item Randomly permute the entries of \(Y\), without replacement within the permutation, to obtain \(\pi_j(Y)\).
  \item Fit \(\widehat{\mathbf f}_{\pi_j(Y)}(\mathbf T)\) and \(\widehat{\mathbf s}_{\pi_j(Y)}(\mathbf T)\).
  \item Calculate \(\mathbf h_{\pi_j(Y)}=\{\widehat{\mathbf f}_{\pi_j(Y)}(\mathbf T)-\overline Y\}/\widehat{\mathbf s}_{\pi_j(Y)}(\mathbf T)\).
  \item Set \(\mathbf M_{\pi(Y)}[j,\cdot]\leftarrow\mathbf h_{\pi_j(Y)}^{\prime}\).
  \end{enumerate}
\item Subtract each column mean of \(\mathbf M_{\pi(Y)}\) and divide by that column's standard deviation to obtain \(\widetilde{\mathbf M}_{\pi(Y)}\).
\item Standardize \(\mathbf M_Y\) using the same column means and standard deviations from \(\mathbf M_{\pi(Y)}\), obtaining \(\widetilde{\mathbf M}_Y\).
\item Calculate the singular value decomposition
\[
 \widetilde{\mathbf M}_{\pi(Y)}=\mathbf U\mathbf\Sigma\mathbf V^{\prime},\qquad
 \mathbf\Sigma=\operatorname{diag}(\sigma_1,\ldots,\sigma_r),\quad
 r=\operatorname{rank}(\widetilde{\mathbf M}_{\pi(Y)}),
\]
with \(\sigma_1\geq\cdots\geq\sigma_r>0\).
\item Calculate the scaled eigenvalues
\[
 \lambda_i=\frac{\sigma_i^2}{\sum_{j=1}^r\sigma_j^2}\,\ensuremath{\mathit{nPoint}},\qquad i=1,\ldots,r.
\]
Their sum is \(\ensuremath{\mathit{nPoint}}\).
\item Let \(k\) be the smallest integer satisfying
\[
 \frac{\sum_{i=1}^k\lambda_i}{\ensuremath{\mathit{nPoint}}}>\frac{\ensuremath{\mathit{percent}}}{100}.
\]
\item Let \(\mathbf V_k\) contain the first \(k\) columns of \(\mathbf V\), and let \(\mathbf\Lambda_k=\operatorname{diag}(\lambda_1,\ldots,\lambda_k)\).
\item Calculate the \(\ensuremath{\mathit{nPerm}}\)-vector of Mahalanobis distances
\[
 \mathbf d_{\pi(Y)}=\sqrt{\operatorname{diag}\!\left(
 \widetilde{\mathbf M}_{\pi(Y)}\mathbf V_k\mathbf\Lambda_k^{-1}\mathbf V_k^{\prime}
 \widetilde{\mathbf M}_{\pi(Y)}^{\prime}\right)}.
\]
\item Calculate the \(\ensuremath{\mathit{nSVEM}}\)-vector of holdout Mahalanobis distances
\[
 \mathbf d_Y=\sqrt{\operatorname{diag}\!\left(
 \widetilde{\mathbf M}_Y\mathbf V_k\mathbf\Lambda_k^{-1}\mathbf V_k^{\prime}
 \widetilde{\mathbf M}_Y^{\prime}\right)}.
\]
\item Fit a SHASH distribution \citep{jones2009sinharcsinh} to \(\mathbf d_{\pi(Y)}\), giving the cumulative distribution function \(F_{\mathrm{SHASH}}\). This empirical choice is not a theoretical null distribution. If SHASH is unavailable, a Weibull or gamma distribution may be substituted.
\item Return the approximate heuristic \(p\)-value
\[
 p=1-\operatorname{median}\!\left\{F_{\mathrm{SHASH}}(\mathbf d_Y)\right\},
\]
where the CDF is applied componentwise.
\item Plot the entries of \(\mathbf d_{\pi(Y)}\) alongside those of \(\mathbf d_Y\). The permutation distances provide an approximate null reference. Under the alternative, the entries of \(\mathbf d_Y\) tend to be larger. The finite-sample calibration of this comparison is assessed empirically.
\end{enumerate}

A scaled beta reference distribution was considered for the squared distances, but simulations indicated inflated Type I error rates in this application. No exact beta distribution is assumed for the holdout distances. Furthermore, a pure randomization test \citep{edgington2007randomization} would generate enough samples from the null distribution to make a direct comparison with the test statistic. However, we make use of the SHASH distribution due to the computational cost of generating the null points: this seems to be a reasonable compromise given the goal of developing a graphical diagnostic that does not require excessive precision in the corresponding p-value.

As constructed, the test assumes that there are no missing values in \(Y\) or missing entries in \(\mathbf{X}\) that could potentially be imputed by the SVEM model, depending on the SVEM programming and modeling framework. If imputation is performed by the SVEM predictor, this could lead to increased error rates (either Type I or Type II are feasible). In cases where either a response or factor setting is missing, it is safer to delete the corresponding rows of \(Y\) and \(\mathbf{X}\) in order to prevent inadvertent imputation. The standard errors of the SVEM predictor can be inflated at the deleted point due to increased prediction variance. As is the case with the ANOVA \emph{F}-test, if the experimental observations are missing completely at random \citep{little2002missing} then the power of the test will decrease due to the smaller sample size. Runs that are missing not at random can lead either to increased Type I or Type II error rates depending on the nature of the bias imparted to the parameter estimates and on the extent of the loss of degrees of freedom for error.

Furthermore, this test could potentially be sensitive to outliers or influential observations. By analogy, large outliers can decrease the power of the \emph{F} test in linear regression by inflating the MSE. The nature and degree of sensitivity of the proposed test to these issues would depend on the base model employed by the SVEM framework. If there are concerns about the influence of any subset of observations, a sensitivity analysis could be performed by running this test with and without the subset included.

\section{Illustrative Example}\label{illustrative-example}

\begin{figure}[htbp]
\centering
\includegraphics[width=0.72\linewidth]{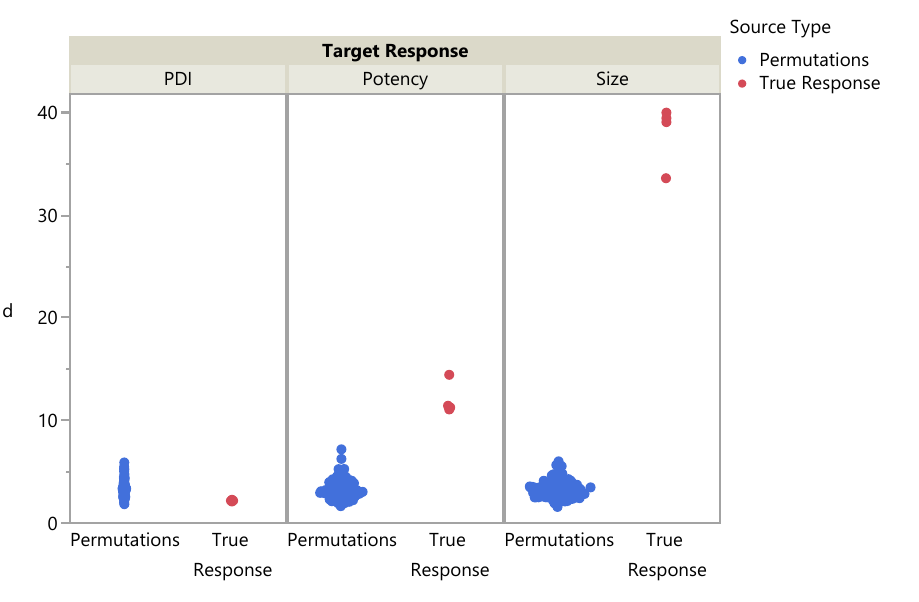}
\caption{Results of the algorithm applied to the SVEM-FS models for the three responses. \(\mathbf{d}_{\pi(Y)}\) is plotted with \(\mathbf{d}_{Y}\). The PDI response was simulated to have no relationship with the study factors. Potency and Size were simulated to have increasingly strong relationships with the study factors.}\label{_Ref147263277}
\end{figure}

\begin{figure}[htbp]
\centering
\includegraphics[width=0.72\linewidth]{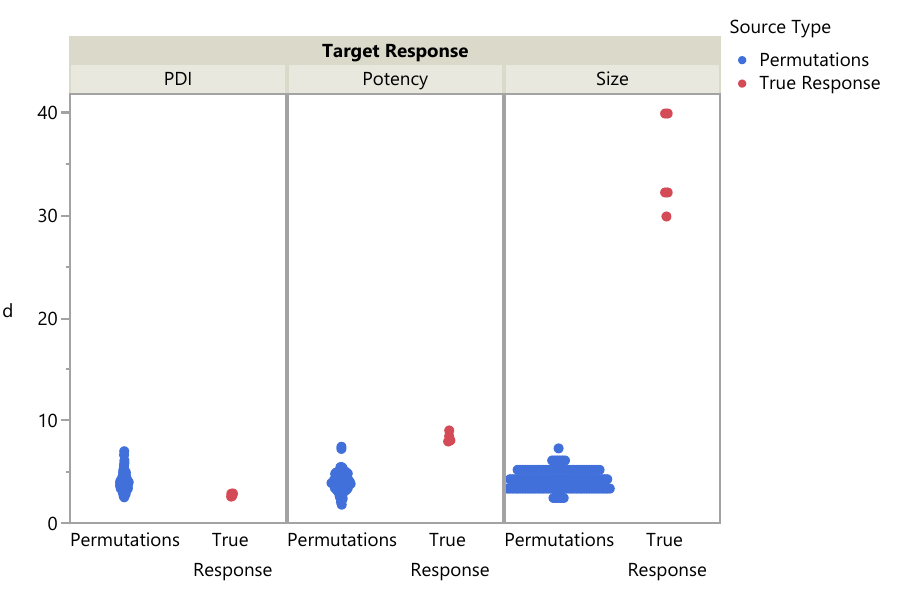}
\caption{Results of the algorithm applied to the SVEM-Lasso models for the three responses. \(\mathbf{d}_{\pi(Y)}\) is plotted with \(\mathbf{d}_{Y}\). The PDI response was simulated to have no relationship with the study factors. Potency and Size were simulated to have increasingly strong relationships with the study factors.}\label{_Ref160795530}
\end{figure}

Returning to the example presented in the Introduction, Table~\ref{_Ref147263156} and Figure~\ref{_Ref147263277} show the results of the test applied to the SVEM-FS model of the simulated experimental results. Table~\ref{_Ref160795520} and Figure~\ref{_Ref160795530} show the results of the test applied to the SVEM-Lasso model. These produce similar conclusions, with both indicating that there is strong evidence that the Potency and Size responses are nonconstant over the range of study factors (which matches the simulation construction). The tests do not reject the null hypothesis for PDI, which was simulated as random noise.

Using these results, the scientist could now consider why a relationship between PDI and the study factors was not detected, and why the relationship for Size is so strong. Section~\ref{limitations-of-the-methodology} provides cautions regarding potential violations of test assumptions that could increase the occurrence of false negatives. As such, the results should be integrated with subject matter expertise as well as other information gleaned from the experimental analysis.

\begin{table}[htbp]
\centering
\caption{Whole-model test results for the three responses modeled with SVEM-FS.}\label{_Ref147263156}
\begin{tabular}{lr}
\toprule
Response & Approximate whole-model $p$-value \\
\midrule
Size & $<0.0001$ \\
Potency & $<0.0001$ \\
PDI & 0.9180 \\
\bottomrule
\end{tabular}
\end{table}

\begin{table}[htbp]
\centering
\caption{Whole-model test results for the three responses modeled with SVEM-Lasso.}\label{_Ref160795520}
\begin{tabular}{lr}
\toprule
Response & Approximate whole-model $p$-value \\
\midrule
Size & $<0.0001$ \\
Potency & 0.0002 \\
PDI & 0.9733 \\
\bottomrule
\end{tabular}
\end{table}

\section{Evaluation of test properties}\label{evaluation-of-test-properties}

Although SVEM is particularly suited for complex models that are built with relatively few observations, we will use a more traditional setting for the power analysis that allows for comparison with the classical ANOVA whole-model test. The design will be a central composite design in three continuous factors ($x_1$, $x_2$, and $x_3$) targeting a RSM as the candidate full model. See \citet{ramsey2023prediction} for a discussion regarding alternative design structures that might be used with SVEM.

\subsection{Power Analysis: Scenario 1}\label{power-analysis-scenario-1}

Scenario 1 simulates values from Equation~\eqref{_Ref155692001}:

\begin{equation}\label{_Ref155692001}
Y = \beta x_{1} + \beta x_{2} + \beta x_{1} x_{2} + \epsilon
\end{equation}
where \(\beta \in \text{[}0,2\text{]}\), and \(\epsilon\sim N(0,1).\) The coefficients for other terms (those involving \(x_{3}\)) in the full response surface candidate model are set to zero in the simulation. The full response surface model (including the terms involving \(x_{3}\)) is used as the candidate set of effects with SVEM-FS and SVEM-Lasso.

\begin{figure}[htbp]
\centering
\includegraphics[width=0.72\linewidth]{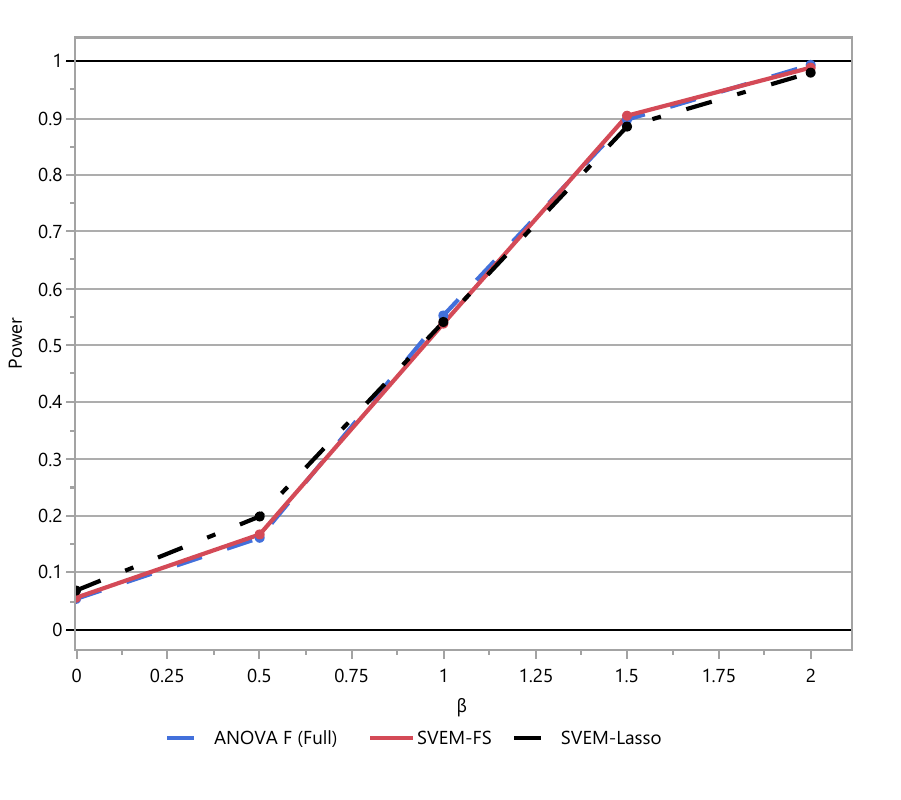}
\caption{Power curves for the ANOVA whole-model test and the proposed test heuristic for SVEM-FS and SVEM-Lasso for Scenario 1.}\label{_Ref147393768}
\end{figure}

Figure~\ref{_Ref147393768} plots the power of the test against \(\beta\), with \(\beta = 0\) representing the null hypothesis. The power analysis is conducted with a nominal Type I error rate of \(\alpha = 0.05\). For comparison, we also record the power of the ANOVA F test for the full response surface model. A total of 1500 simulations are run per model per knot.

In this scenario, the proposed test approximates the nominal Type I error rate for both SVEM-FS and SVEM-Lasso. The power curve for the ANOVA whole-model F-test for the full response surface model is nearly identical to those produced by SVEM-FS and SVEM-Lasso.

\subsection{Power Analysis: Scenario 2}\label{power-analysis-scenario-2}

Scenario 2 is identical to Scenario 1, except only the true reduced model (Equation~\eqref{_Ref155692001}) is used as the candidate set for SVEM-FS and SVEM-Lasso. This is compared to the ANOVA F test for the true reduced model.

\begin{figure}[htbp]
\centering
\includegraphics[width=0.72\linewidth]{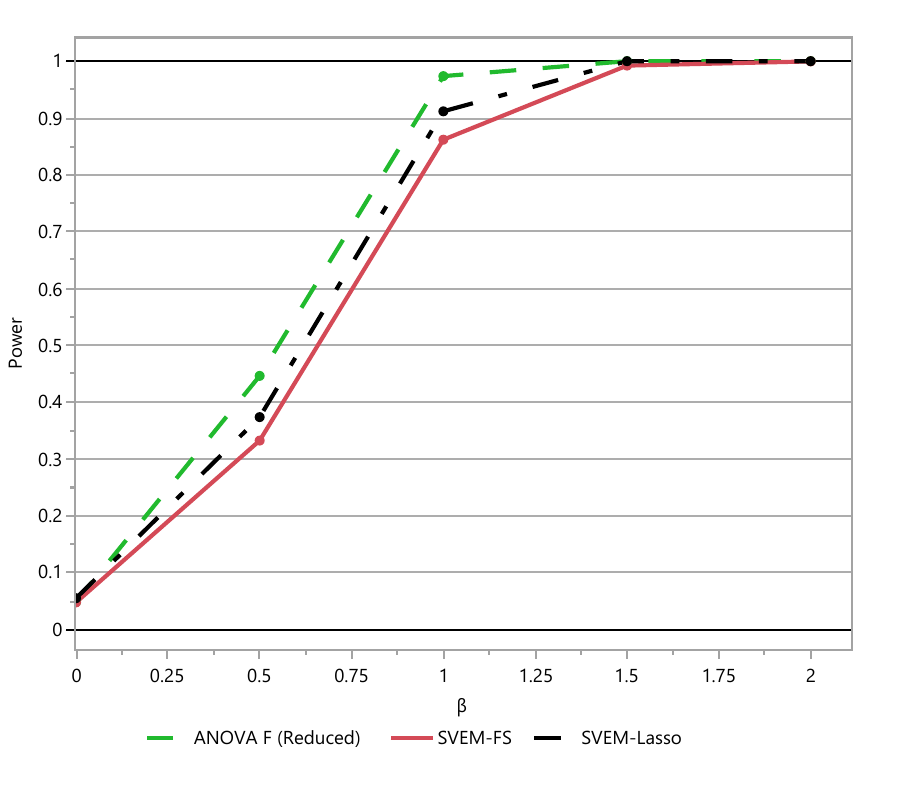}
\caption{Power curves for the ANOVA whole-model test and the proposed test heuristic for SVEM-FS and SVEM-Lasso for Scenario 2.}\label{_Ref155669799}
\end{figure}

Figure~\ref{_Ref155669799} shows that the observed Type I error rates for SVEM-FS and SVEM-Lasso are close to the nominal level. Both SVEM-FS and SVEM-Lasso are less powerful than the ANOVA F test for the reduced model, and SVEM-Lasso appears to be slightly more powerful than SVEM-FS. This simulation represents an unlikely situation in which the candidate model contains no irrelevant parameters. \citet{lemkus2021svem} refer to this as a ``no sparsity'' scenario.

\subsection{Power Analysis: Scenario 3}\label{power-analysis-scenario-3}

Scenario 3 simulates values from Equation~\eqref{_Ref155692030}

\begin{equation}\label{_Ref155692030}
Y = \beta x_{3} + \epsilon
\end{equation}
where \(\beta \in \text{[}0,2\text{]}\), and \(\epsilon\sim N(0,1).\) The coefficients for other terms in the full response surface candidate model are set to zero in the simulation. In this scenario, the active effects are more sparse in the candidate full response surface model than in Scenario 1. The permutation test for both SVEM-FS and SVEM-Lasso when targeting the full response surface is compared to two ANOVA F tests: one corresponding to the full response surface model (including the terms whose coefficients are set to zero in the simulation) and a second corresponding to the reduced model (Equation~\eqref{_Ref155692030}).

\begin{figure}[htbp]
\centering
\includegraphics[width=0.72\linewidth]{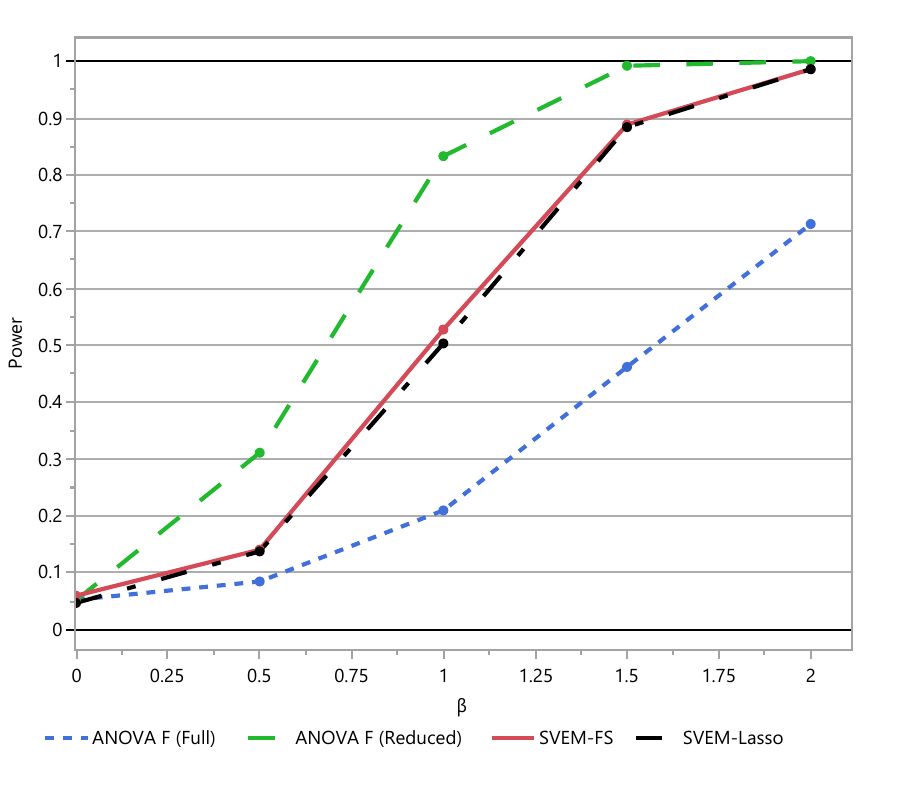}
\caption{Power curves for the ANOVA whole-model test and the proposed test heuristic for SVEM-FS and SVEM-Lasso for Scenario 3.}\label{_Ref155671056}
\end{figure}

Figure~\ref{_Ref155671056} shows that SVEM-FS and SVEM-Lasso yield nearly identical results. These are more powerful than the corresponding ANOVA F test for the full model, and less powerful than the ANOVA F test for the true reduced model. It is encouraging that -- for larger values of \(\beta\) -- the power of SVEM (-FS and -Lasso) is closer to that of the reduced model than the full model, even though SVEM is also using the full model as its base model.

\subsection{Further Graphical Exploration of Scenario 1}\label{further-graphical-exploration-of-scenario-1}

The predictions of the SVEM ensemble contained in \(\mathbf{P}_{\left( \mathbf{X},Y \right)}\left( \mathbf{T} \right)\) allow visualization of the response surface across the factor space with appropriate sampling variation included. This can be a useful summary graphic during an experimental analysis. For Scenario 1, generate a \(50,000 \times 3\) matrix \(\mathbf{T}\) of points randomly spaced throughout the factor region displayed below. Using SVEM-FS from the simulations of Scenario 1, calculate a vector \(\mathbf{v} = \operatorname{RandomNormal}\left( \widehat{\mathbf{f}}_{Y}\left( \mathbf{T} \right),\widehat{\mathbf{s}}_{Y}\left( \mathbf{T} \right) \right)\) for a SVEM fit for a single simulation for each of the three values of \(\beta.\)

\begin{figure}[htbp]
\centering
\includegraphics[width=0.68\linewidth]{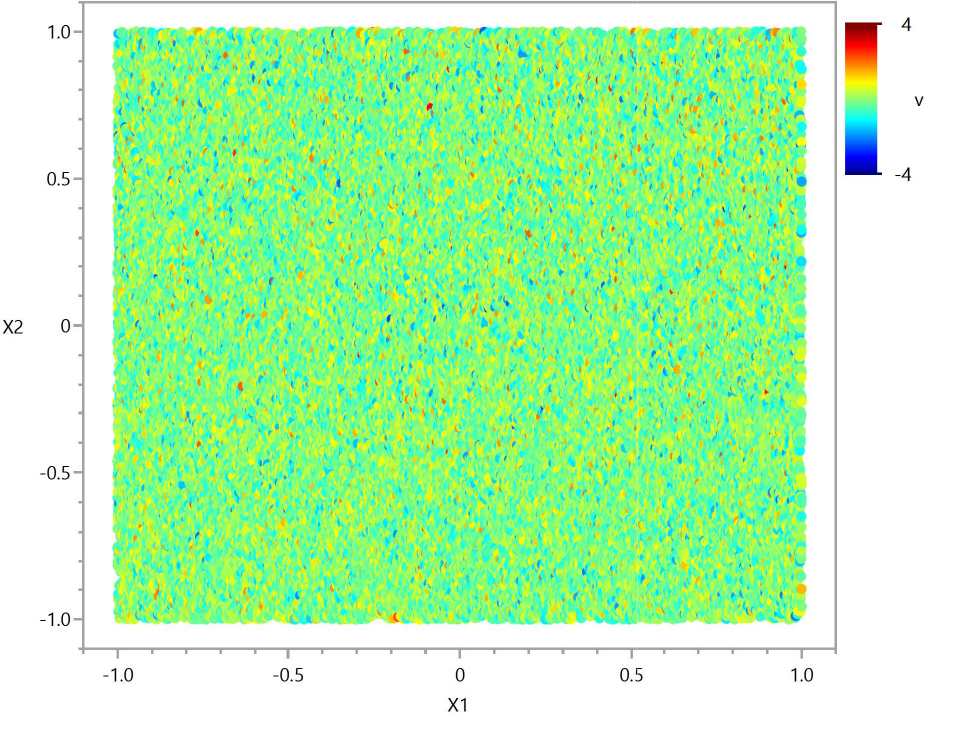}
\caption{Sampled response surface for one simulation for SVEM-FS for $\beta=0$.}\label{_Ref148998273}
\end{figure}

Figure~\ref{_Ref148998273} shows the sampled fitted response surface over \(x_{1}\) and \(x_{2}\) in a single simulation when \(\beta = 0\). The points are colored by the value of \(\mathbf{v}\). A fitted response surface with no structure will be deemed significant by the proposed test in approximately 5\% of attempts (the nominal Type I error rate).

\begin{figure}[htbp]
\centering
\includegraphics[width=0.68\linewidth]{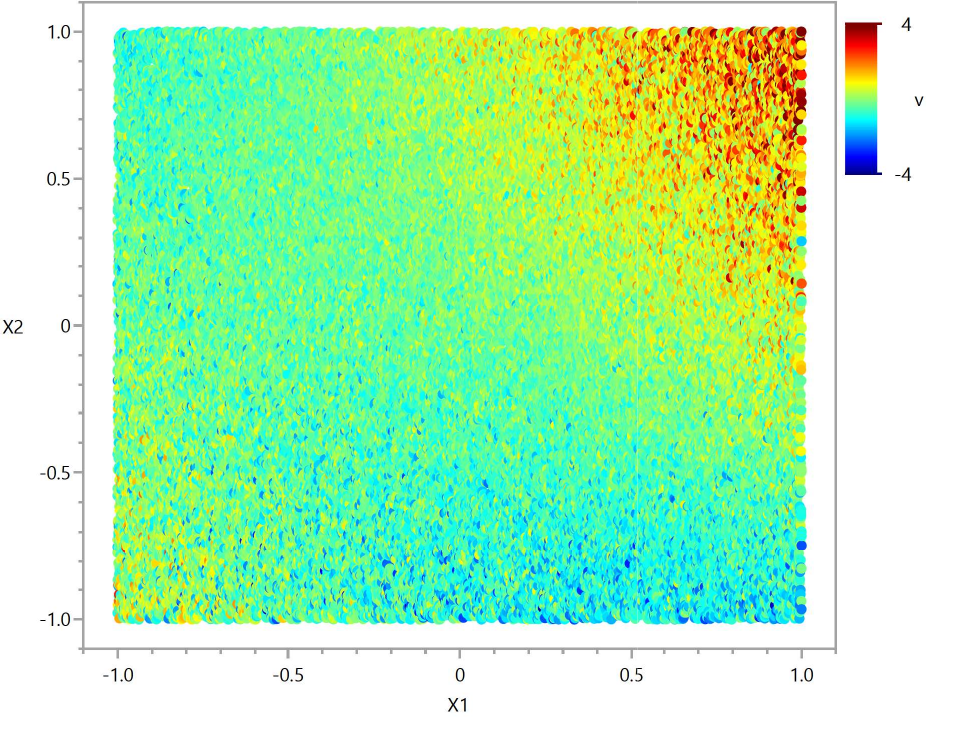}
\caption{Sampled response surface for one simulation for SVEM-FS for $\beta=1.0$.}\label{_Ref148998667}\label{_Ref161186361}
\end{figure}

Figure~\ref{_Ref148998667} shows the sampled response surface over \(x_{1}\) and \(x_{2}\) in the simulation when \(\beta = 1\). A fitted response surface with this amount of structure will be deemed significant by the test in about 55\% of attempts according to the power curve in Figure~\ref{_Ref147393768}. To belabor the point, this graph helps to illustrate the risk of a false negative test result (either from the SVEM permutation test or the ANOVA \emph{F} (full model) test), as these would fail to detect the nonconstant response surface structure shown in Figure~\ref{_Ref161186361} in about 45\% of repeated experiments of the same size.

\begin{figure}[htbp]
\centering
\includegraphics[width=0.68\linewidth]{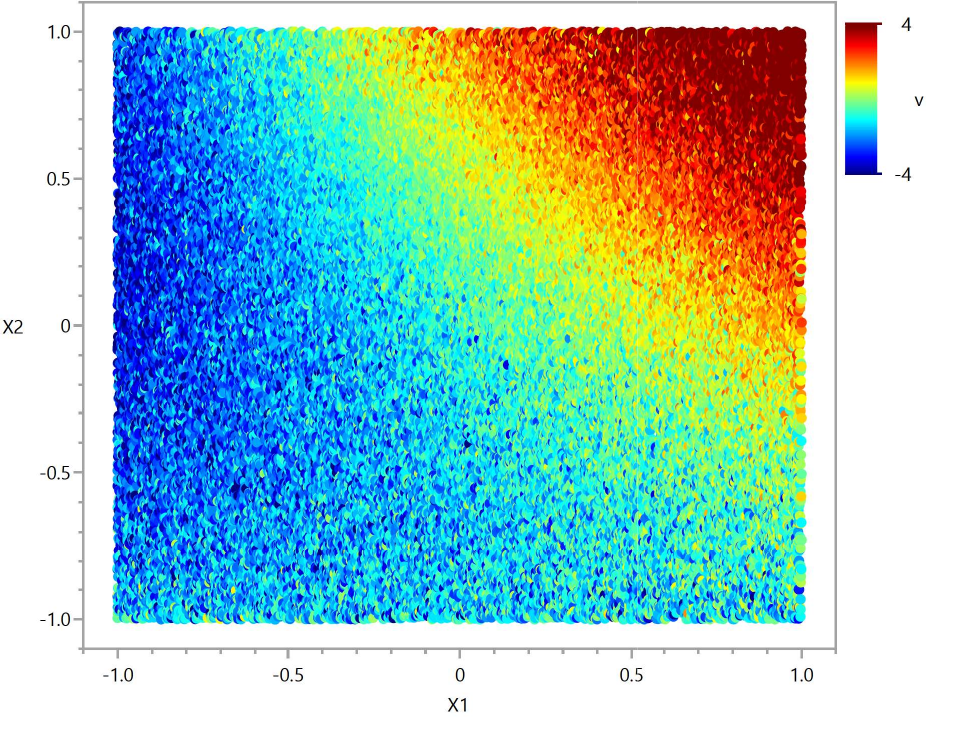}
\caption{Sampled response surface for one simulation for SVEM-FS for \(\beta = 1.75\).}\label{_Ref148999166}
\end{figure}

Figure~\ref{_Ref148999166} shows the sampled response surface over \(x_{1}\) and \(x_{2}\) in the simulation when \(\beta = 1.75\). A fitted response surface with this amount of structure will be deemed significant by the test in about 95\% of attempts, according to the power curve in Figure~\ref{_Ref147393768}.

\section{Limitations of the Methodology}\label{limitations-of-the-methodology}

The effectiveness of the test is contingent upon the chosen model's ability to represent the true response surface as well as the adequacy of the study design. The following circumstances can increase the probability of a Type II error:

\begin{itemize}
\item
  Insufficient study size relative to the measurement and process variance. Larger experimental studies are needed for ``noisy'' processes. In such systems, if the design is too small it is possible that this test will fail to detect a relationship between the response and the study factors even though there may be a deterministic physical relationship that is adequately described within the candidate model structure used by SVEM.
\item
  The inclusion of too many insignificant study factors, or an overly complex target model for SVEM. This is demonstrated in the comparison of the SVEM power in Figure~\ref{_Ref147393768} (Scenario 1) to the increased SVEM power in Figure~\ref{_Ref155669799} (Scenario 2) when the irrelevant parameters have been removed from the base model used by SVEM.
\item
  The model targeted by SVEM is biased (too rigid relative to the complexity of the true response surface). The SVEM bootstrap reduces model variance by averaging over potentially high-variance predictions. However, if the base model used with SVEM is biased, then the resulting SVEM predictions will also be biased. Adding effects to the candidate set considered by SVEM (any important covariates that had been omitted, including simple additive blocking factors), or employing more flexible base models such as boosted trees or neural networks might increase the power of the test.
\item
  The study design does not sufficiently capture areas where the true response surface changes. For instance, a response surface that is flat at the boundaries but exhibits a peak (or valley) in the middle could be misrepresented in models based solely on boundary runs.
\end{itemize}

If a boundary-heavy design is employed, there can be regions of large prediction variance on the interior of the factor space, even when the correct model is specified during design construction. Since this test relies on the \emph{nPoint} locations spaced throughout the factor space, it is possible that this test could be underpowered in these scenarios. Future research could explore different design configurations and levels of complexity in the response surface. \citet{ramsey2023prediction} advocate including test points in an experiment in order to evaluate the fit of the SVEM models and to reduce the prediction variance in these interior regions. A reduction in interior prediction variance could be beneficial for optimization applications, even beyond the scope of assessing model fit and aiding the whole-model test.

\section{Final Comments}\label{final-comments}

The proposed test heuristic provides a tool to test for nonconstancy in the modeled SVEM response surface and therefore to help identify spurious correlation between predicted and actual response values. A side-by-side comparison of the resulting distance plots for each of the study responses could be informative for subject matter experts: perhaps one of the responses should receive a larger sample size in the future due to greater measurement variance, or perhaps a more complex study design may be suggested. When setting importance weights for joint optimization of study responses, consider if responses that show a stronger significance of fit should receive relatively larger weights than subject-matter importance might otherwise impose; carefully prioritizing responses with stronger evidence of relationship to study factors could enhance the optimization process by reducing the impact of responses with weaker correlations.

The power of the proposed test depends on the suitability of the employed experimental design in addition to properties of the selected SVEM framework. It is ideal to build interior test points into an experimental design in order to have an independent test set to assess the quality of the resulting model fit (SVEM or otherwise) and to reduce prediction variance on the interior of the factor space once these points are included in a final model (allowing SVEM to use all observations, including the test set). When a test set is not available, the proposed test can provide an internal measure of model fit.

Finally, although SVEM is used for predictive purposes rather than studying significance of individual parameters or effects, a similar randomization approach could potentially be used to develop a variable importance diagnostic by permuting the factor columns one at a time instead of permuting the response column. Alternatively, the Bayesian Probability of Agreement method \citep{stevens2020agreement} could feasibly be used to compare the equivalence of response surfaces arising with and without certain factors included.

\section*{Acknowledgements}

Thank you to Clay Barker, Marie Gaudard, Chris Gotwalt, and Phil Ramsey for patiently answering questions regarding SVEM over the past several years. Marie caught errors in an early draft of this paper (any current errors are my own). Thank you to the editor and referees for helpful corrections and suggestions.

\appendix
\section{Illustrative-example model specification}\label{appendix-a}

The 100 candidate fixed effect parameters used by SVEM-FS and SVEM-Lasso in the illustrative example are listed below. An intercept is included in the model and mixture main effects are not forced \citep{karl2022mixtures,karl2023lnp}. Table~\ref{_Ref160927988} lists the study factors and their ranges.

\begin{table}[htbp]
\centering
\caption{Study factors and ranges.}\label{_Ref160927988}
\begin{tabular}{lll}
\toprule
Factor name & Role & Values \\
\midrule
PEG & Mixture & $[0.01,0.05]$ \\
Helper & Mixture & $[0.1,0.6]$ \\
Ionizable & Mixture & $[0.1,0.6]$ \\
Cholesterol & Mixture & $[0.1,0.6]$ \\
Ionizable Lipid Type & Categorical & $\{\mathrm{H101},\mathrm{H102},\mathrm{H103}\}$ \\
\texttt{N\_P\_ratio} & Continuous & $[6,14]$ \\
flow rate & Continuous & $[1,3]$ \\
\bottomrule
\end{tabular}
\end{table}

The 76 effect expressions below include 24 containing the three-level categorical factor. Each such expression contributes two coefficient columns, giving 100 candidate coefficient columns in total.

\begin{longtable}{r p{0.88\linewidth}}
\toprule
No. & Model effect (original JMP notation) \\
\midrule
\endfirsthead
\toprule No. & Model effect (continued) \\ \midrule
\endhead
\bottomrule\endlastfoot
1 & \texttt{\detokenize{(Intercept)}} \\
2 & \texttt{\detokenize{:PEG & Mixture}} \\
3 & \texttt{\detokenize{:Helper & Mixture}} \\
4 & \texttt{\detokenize{:Ionizable & Mixture}} \\
5 & \texttt{\detokenize{:Cholesterol & Mixture}} \\
6 & \texttt{\detokenize{:Ionizable Lipid Type}} \\
7 & \texttt{\detokenize{:N_P_ratio & RS}} \\
8 & \texttt{\detokenize{:flow rate & RS}} \\
9 & \texttt{\detokenize{:PEG * :Helper}} \\
10 & \texttt{\detokenize{:PEG * :Ionizable}} \\
11 & \texttt{\detokenize{:Helper * :Ionizable}} \\
12 & \texttt{\detokenize{:PEG * :Cholesterol}} \\
13 & \texttt{\detokenize{:Helper * :Cholesterol}} \\
14 & \texttt{\detokenize{:Ionizable * :Cholesterol}} \\
15 & \texttt{\detokenize{Scheffe Cubic( :PEG, :Helper )}} \\
16 & \texttt{\detokenize{Scheffe Cubic( :PEG, :Ionizable )}} \\
17 & \texttt{\detokenize{:PEG * :Helper * :Ionizable}} \\
18 & \texttt{\detokenize{Scheffe Cubic( :Helper, :Ionizable )}} \\
19 & \texttt{\detokenize{Scheffe Cubic( :PEG, :Cholesterol )}} \\
20 & \texttt{\detokenize{:PEG * :Helper * :Cholesterol}} \\
21 & \texttt{\detokenize{Scheffe Cubic( :Helper, :Cholesterol )}} \\
22 & \texttt{\detokenize{:PEG * :Ionizable * :Cholesterol}} \\
23 & \texttt{\detokenize{:Helper * :Ionizable * :Cholesterol}} \\
24 & \texttt{\detokenize{Scheffe Cubic( :Ionizable, :Cholesterol )}} \\
25 & \texttt{\detokenize{:Ionizable Lipid Type * :N_P_ratio}} \\
26 & \texttt{\detokenize{:N_P_ratio * :N_P_ratio}} \\
27 & \texttt{\detokenize{:Ionizable Lipid Type * :flow rate}} \\
28 & \texttt{\detokenize{:N_P_ratio * :flow rate}} \\
29 & \texttt{\detokenize{:flow rate * :flow rate}} \\
30 & \texttt{\detokenize{:N_P_ratio * :N_P_ratio * :Ionizable Lipid Type}} \\
31 & \texttt{\detokenize{:N_P_ratio * :N_P_ratio * :flow rate}} \\
32 & \texttt{\detokenize{:flow rate * :flow rate * :Ionizable Lipid Type}} \\
33 & \texttt{\detokenize{:flow rate * :flow rate * :N_P_ratio}} \\
34 & \texttt{\detokenize{:PEG * :Ionizable Lipid Type}} \\
35 & \texttt{\detokenize{:PEG * :N_P_ratio}} \\
36 & \texttt{\detokenize{:PEG * :flow rate}} \\
37 & \texttt{\detokenize{:Helper * :Ionizable Lipid Type}} \\
38 & \texttt{\detokenize{:Helper * :N_P_ratio}} \\
39 & \texttt{\detokenize{:Helper * :flow rate}} \\
40 & \texttt{\detokenize{:Ionizable * :Ionizable Lipid Type}} \\
41 & \texttt{\detokenize{:Ionizable * :N_P_ratio}} \\
42 & \texttt{\detokenize{:Ionizable * :flow rate}} \\
43 & \texttt{\detokenize{:Cholesterol * :Ionizable Lipid Type}} \\
44 & \texttt{\detokenize{:Cholesterol * :N_P_ratio}} \\
45 & \texttt{\detokenize{:Cholesterol * :flow rate}} \\
46 & \texttt{\detokenize{:PEG * :Helper * :Ionizable Lipid Type}} \\
47 & \texttt{\detokenize{:PEG * :Helper * :N_P_ratio}} \\
48 & \texttt{\detokenize{:PEG * :Helper * :flow rate}} \\
49 & \texttt{\detokenize{:PEG * :Ionizable * :Ionizable Lipid Type}} \\
50 & \texttt{\detokenize{:PEG * :Ionizable * :N_P_ratio}} \\
51 & \texttt{\detokenize{:PEG * :Ionizable * :flow rate}} \\
52 & \texttt{\detokenize{:PEG * :Cholesterol * :Ionizable Lipid Type}} \\
53 & \texttt{\detokenize{:PEG * :Cholesterol * :N_P_ratio}} \\
54 & \texttt{\detokenize{:PEG * :Cholesterol * :flow rate}} \\
55 & \texttt{\detokenize{:PEG * :Ionizable Lipid Type * :N_P_ratio}} \\
56 & \texttt{\detokenize{:PEG * :Ionizable Lipid Type * :flow rate}} \\
57 & \texttt{\detokenize{:PEG * :N_P_ratio * :flow rate}} \\
58 & \texttt{\detokenize{:Helper * :Ionizable * :Ionizable Lipid Type}} \\
59 & \texttt{\detokenize{:Helper * :Ionizable * :N_P_ratio}} \\
60 & \texttt{\detokenize{:Helper * :Ionizable * :flow rate}} \\
61 & \texttt{\detokenize{:Helper * :Cholesterol * :Ionizable Lipid Type}} \\
62 & \texttt{\detokenize{:Helper * :Cholesterol * :N_P_ratio}} \\
63 & \texttt{\detokenize{:Helper * :Cholesterol * :flow rate}} \\
64 & \texttt{\detokenize{:Helper * :Ionizable Lipid Type * :N_P_ratio}} \\
65 & \texttt{\detokenize{:Helper * :Ionizable Lipid Type * :flow rate}} \\
66 & \texttt{\detokenize{:Helper * :N_P_ratio * :flow rate}} \\
67 & \texttt{\detokenize{:Ionizable * :Cholesterol * :Ionizable Lipid Type}} \\
68 & \texttt{\detokenize{:Ionizable * :Cholesterol * :N_P_ratio}} \\
69 & \texttt{\detokenize{:Ionizable * :Cholesterol * :flow rate}} \\
70 & \texttt{\detokenize{:Ionizable * :Ionizable Lipid Type * :N_P_ratio}} \\
71 & \texttt{\detokenize{:Ionizable * :Ionizable Lipid Type * :flow rate}} \\
72 & \texttt{\detokenize{:Ionizable * :N_P_ratio * :flow rate}} \\
73 & \texttt{\detokenize{:Cholesterol * :Ionizable Lipid Type * :N_P_ratio}} \\
74 & \texttt{\detokenize{:Cholesterol * :Ionizable Lipid Type * :flow rate}} \\
75 & \texttt{\detokenize{:Cholesterol * :N_P_ratio * :flow rate}} \\
76 & \texttt{\detokenize{:Ionizable Lipid Type * :N_P_ratio * :flow rate}} \\
\end{longtable}
\addtocounter{table}{-1}

\section{Choice of test parameters}\label{appendix-b}

This appendix discusses the default parameter settings shown in Table~\ref{_Ref160928120}.

\begin{table}[htbp]
\centering
\caption{Default test parameter settings.}\label{_Ref160928120}
\begin{tabular}{llp{0.63\linewidth}}
\toprule
Setting & Default & Note \\
\midrule
\textit{nPerm} & 125 & Number of permutations $\pi(Y)$ fit with SVEM \\
\textit{nPoint} & 2000 & Number of points sampled throughout factor space \\
\textit{nBoot} & 200 & Number of internal SVEM bootstraps per fit \\
\textit{percent} & 85 & Threshold for the cumulative percentage of the eigenvalue sum (Step 11) \\
\textit{nSVEM} & 5 & Number of times SVEM is refit to $Y$ \\
\bottomrule
\end{tabular}
\end{table}

\begin{figure}[htbp]
\centering
\includegraphics[width=0.72\linewidth]{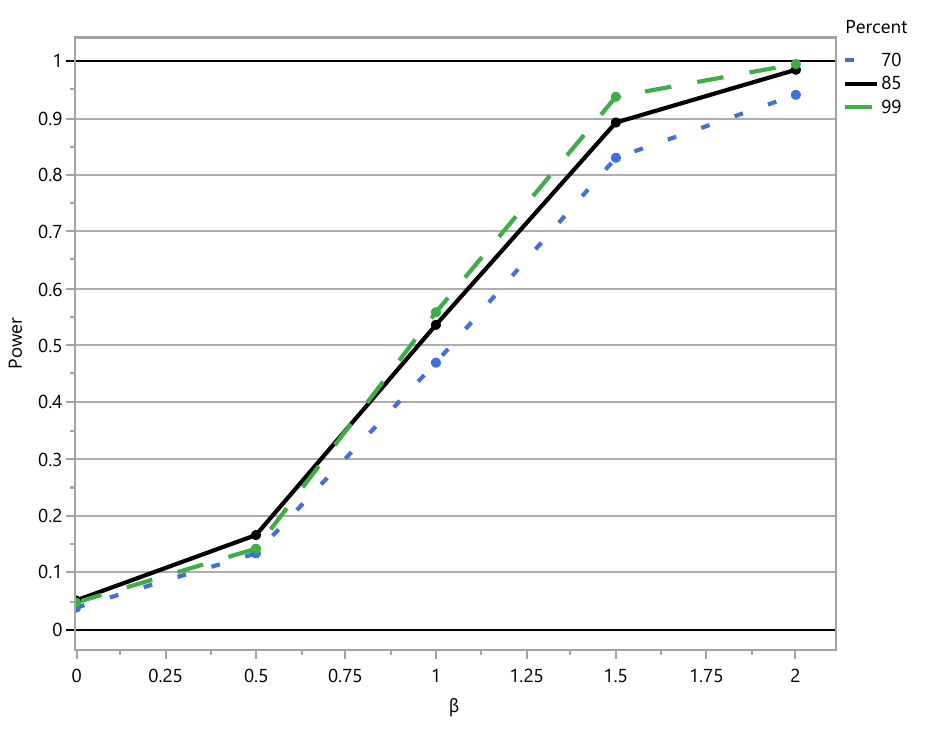}
\caption{Power analysis for Scenario 1 using default settings (\ensuremath{\mathit{percent}}=85, black line) compared with two alternate settings for \ensuremath{\mathit{percent}}.}\label{_Ref161179046}
\end{figure}

\begin{itemize}
\item
  The \emph{percent} parameter controls the rank reduction of the SVD used to calculate the Mahalanobis distance. Figure~\ref{_Ref161179046} shows the results of changing \emph{percent} to 70 and 99 in Scenario 1, compared with the default of 85. Although it did not occur in this application, setting \emph{percent} to a high value such as 99\% could potentially inflate the Type I error rate above nominal. Figure~\ref{_Ref161179046} illustrates how excessive reduction can dampen the power of the test.
\end{itemize}

\begin{figure}[htbp]
\centering
\includegraphics[width=0.68\linewidth]{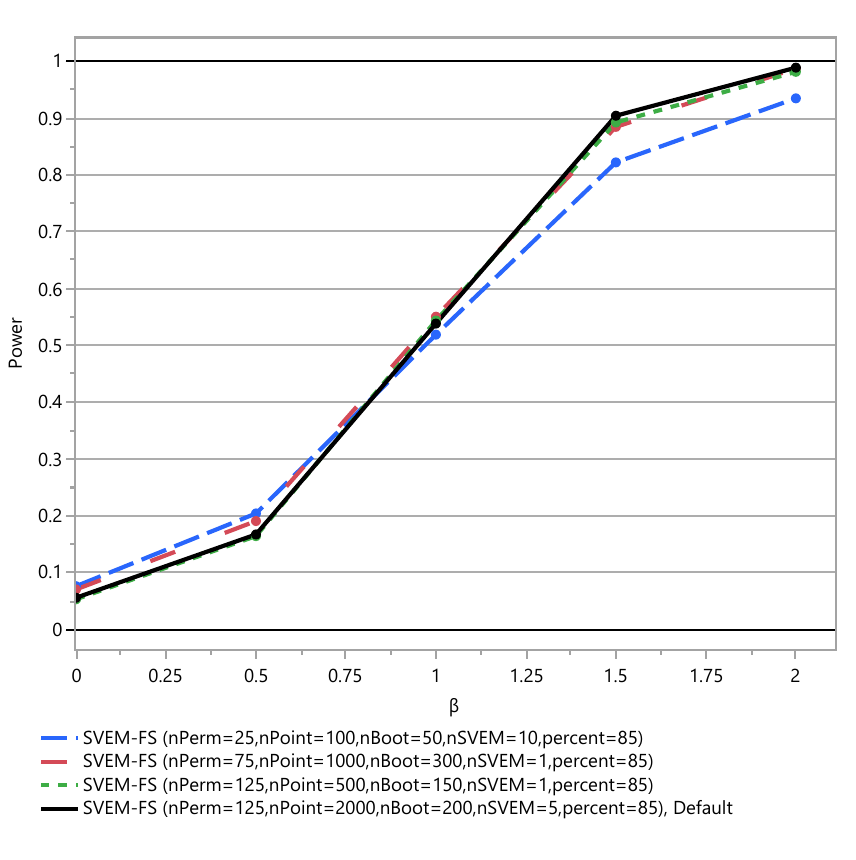}
\caption{Power analysis for Scenario 1 using default settings (black line) compared with three alternate settings for \ensuremath{\mathit{nPerm}}, \ensuremath{\mathit{nPoint}}, \ensuremath{\mathit{nBoot}}, and \ensuremath{\mathit{nSVEM}}.}\label{_Ref161178702}
\end{figure}

\begin{figure}[htbp]
\centering
\includegraphics[width=0.68\linewidth]{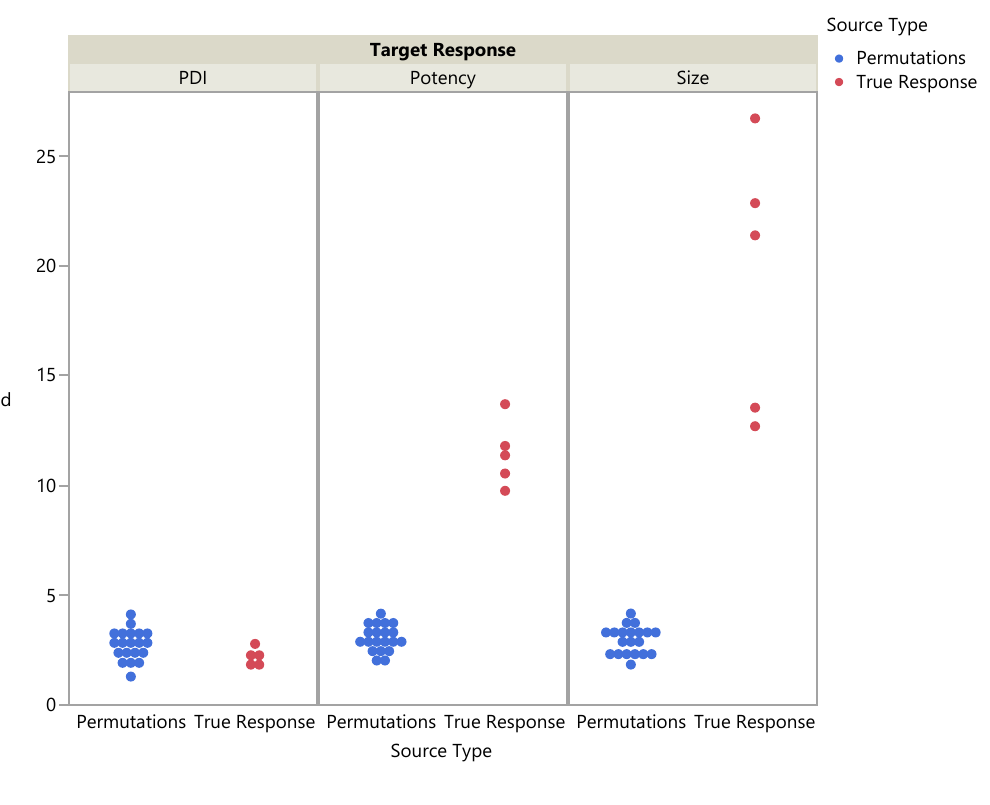}
\caption{Results of the test with alternate settings (\ensuremath{\mathit{nPerm}}=20, \ensuremath{\mathit{nPoint}}=100, \ensuremath{\mathit{nBoot}}=30, \ensuremath{\mathit{percent}}=85, \ensuremath{\mathit{nSVEM}}=5) applied to the SVEM-FS models for PDI, Potency, and Size.}\label{_Ref161070435}
\end{figure}

\begin{itemize}
\item
  The choice of \emph{nPerm=125} is motivated by the desire to capture a p-value of about 0.01 with the permutation reference distribution alone. Fitting the 125 values with a SHASH distribution allows us to increase the granularity of the p-value by substituting a flexible parametric structure for larger values of \emph{nPerm}.
\item
  \emph{nPoint} should be large enough to cover the factor space at sufficient density to capture regions of the response surface that differ from its average value. The setting of 2000 is likely excessive for these applications; smaller values produced similar results (Figure~\ref{_Ref161178702} and Figure~\ref{_Ref161070435}).
\item
  \emph{nBoot} is set to the default of 200 recommended by \citet{lemkus2021svem} and should match the setting used by SVEM in the application of interest.
\item
  \emph{nSVEM} is set to a value greater than 1 in order to include a graphical summary of the bootstrap variance of the SVEM estimates. Alternatives to using the median of the \emph{nSVEM} distances in Algorithm~\ref{alg:whole-model} would be either to simply use a single model fit to the unpermuted response vector or to report all \emph{nSVEM} p-values to capture the impact of the bootstrap variance on the result.
\end{itemize}

Figure~\ref{_Ref161178702} compares the power curves for Scenario 1 using different settings for these parameters. Most of the curves nearly overlap, except one combination corresponding to particularly small values of \(\mathit{nPerm}=25\), \(\mathit{nPoint}=100\), and \(\mathit{nBoot}=50\), with \(\mathit{nSVEM}=10\).

\begin{figure}[htbp]
\centering
\includegraphics[width=0.68\linewidth]{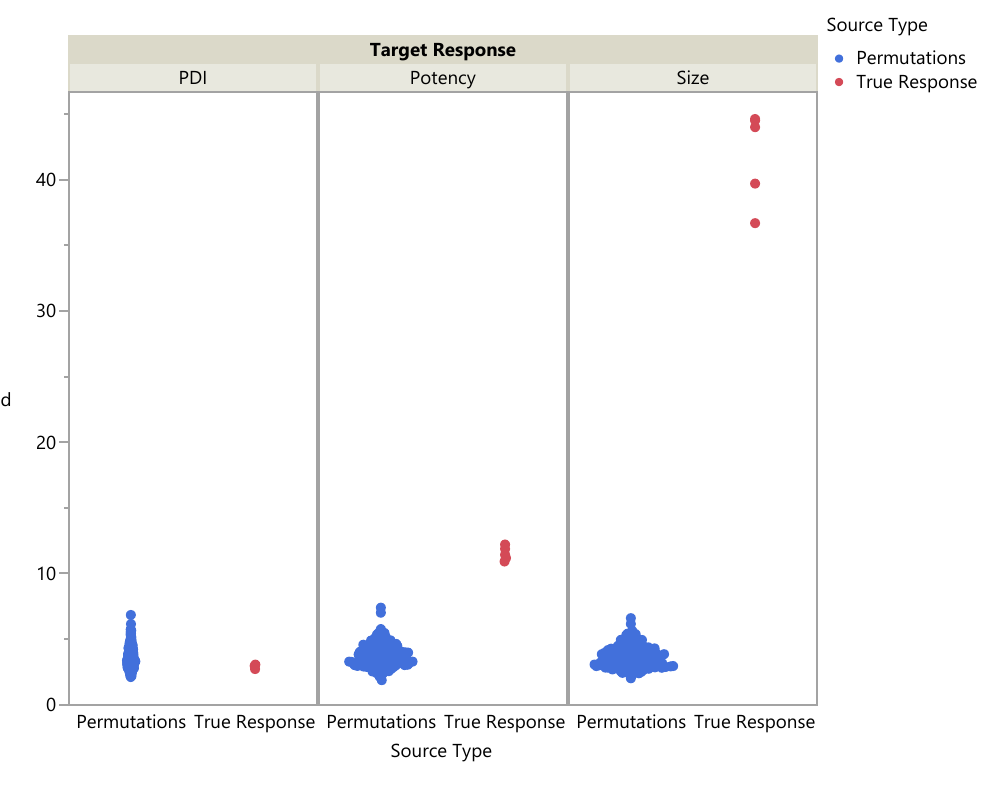}
\caption{Results of the test with alternate settings (\ensuremath{\mathit{nPerm}}=200, \ensuremath{\mathit{nPoint}}=3000, \ensuremath{\mathit{nBoot}}=500, \ensuremath{\mathit{percent}}=85, \ensuremath{\mathit{nSVEM}}=5) applied to the SVEM-FS models for PDI, Potency, and Size.}\label{_Ref161178997}
\end{figure}

The supplementary document \textbf{LNP\_test\_results.pdf} contains the test results and associated graphics for 14 different combinations of parameter settings for the LNP example using SVEM-FS. Two additional graphs are presented here: one using less computationally demanding settings than the default, and the second using more demanding settings. Figure~\ref{_Ref161070435} and Table~\ref{_Ref161070470} report the application of the permutation test to the LNP example using \emph{nPerm}=20, \emph{nPoint}=100, \emph{nBoot}=30, \emph{percent}=85, and \emph{nSVEM}=5. These are markedly smaller than the default settings yet reach the same general conclusion as the application of more demanding settings (\emph{nPerm}=200, \emph{nPoint}=3000, \emph{nBoot}=500, \emph{percent}=85, and \emph{nSVEM}=5) shown in Figure~\ref{_Ref161178997} and Table~\ref{_Ref161179007}.

\begin{table}[htbp]
\centering
\caption{Approximate whole-model $p$-values with alternate settings ($\ensuremath{\mathit{nPerm}}=20$, $\ensuremath{\mathit{nPoint}}=100$, $\ensuremath{\mathit{nBoot}}=30$, $\ensuremath{\mathit{percent}}=85$, $\ensuremath{\mathit{nSVEM}}=5$).}\label{_Ref161070470}
\begin{tabular}{lr}
\toprule
Response & Approximate whole-model $p$-value \\
\midrule
Potency & $<0.0001$ \\
Size & $<0.0001$ \\
PDI & 0.8410 \\
\bottomrule
\end{tabular}
\end{table}

\begin{table}[htbp]
\centering
\caption{Approximate whole-model $p$-values with alternate settings ($\ensuremath{\mathit{nPerm}}=200$, $\ensuremath{\mathit{nPoint}}=3000$, $\ensuremath{\mathit{nBoot}}=500$, $\ensuremath{\mathit{percent}}=85$, $\ensuremath{\mathit{nSVEM}}=5$).}\label{_Ref161179007}
\begin{tabular}{lr}
\toprule
Response & Approximate whole-model $p$-value \\
\midrule
Size & $<0.0001$ \\
Potency & $<0.0001$ \\
PDI & 0.7328 \\
\bottomrule
\end{tabular}
\end{table}

\section*{Supplemental material}\label{supplemental-material}

The accompanying ancillary directory contains the following files. The supplied code and numerical results are retained from the original study.

\begin{itemize}
\item
  \texttt{JSL\_code\_distance\_calc.jsl}: JMP code for Algorithm~\ref{alg:whole-model}, starting with Step 7, using a SHASH reference fit. The R example implements the same distance calculations with a Weibull reference fit.
\item
  \texttt{LNP\_test\_results.pdf}: Results of the application of the permutation test for SVEM-FS for the LNP optimization example using 14 different combinations of test parameter settings.
\item
  \texttt{M\_pi\_Y.csv}: CSV input for the R and JSL distance-calculation examples. This is an example matrix produced by Steps 1--6 of Algorithm~\ref{alg:whole-model}.
\item
  \texttt{M\_Y.csv}: CSV input for the R and JSL distance-calculation examples. This is an example matrix produced by Steps 1--6 of Algorithm~\ref{alg:whole-model}.
\item
  \texttt{R\_code\_distance\_calc.R}: R code for Algorithm~\ref{alg:whole-model}, starting with Step 7. It uses a Weibull reference fit instead of SHASH, so its final approximate $p$-value can differ from that of the JSL example.
\item
  \texttt{simulated\_LNP\_example.csv}: CSV file for the example in the Introduction
\item
  \texttt{simulated\_LNP\_example.jmp}: JMP data table for the example in the Introduction
\item
  \texttt{SVEM\_whole\_model\_v7.1\_jsl.txt}: A text file containing a JMP 17 Pro script that runs the proposed test
\end{itemize}

\begingroup\raggedright
\nocite{karl2015text}
\bibliographystyle{plainnat}
\bibliography{references}
\endgroup
\end{document}